\documentclass[12pt]{article}
\usepackage{stmaryrd}
\usepackage{amssymb}
\usepackage{graphics}
\usepackage{epsfig}
\usepackage{a4wide}
\usepackage{cite}

\begin{document}
\newcommand{\eqehq}{$e^-q \to e^-h^0q$ }
\newcommand{\eqehqg}{$e^-q \to e^-h^0qg$ }
\newcommand{\egehqq}{$e^-g \to e^-h^0q\bar q$ }
\newcommand{\epehj}{$e^-p \to e^-h^0j+X$ }

\title{ The light MSSM neutral Higgs boson production
associated with an electron and a jet at the LHeC  }
\author{ Wen Zhe, Wang Shao-Ming, Ma Wen-Gan, Guo Lei, and Zhang Ren-You \\
{\small Department of Modern Physics, University of Science and Technology}  \\
{\small of China (USTC), Hefei, Anhui 230026, People's Republic of China}  }

\date{}
\maketitle \vskip 15mm
\begin{abstract}
We study the light $CP$-even neutral Higgs boson production in
association with an electron and a jet at the possible CERN large
hadron-electron collider within the minimal supersymmetric
standard model. We investigate the possible supersymmetric
effects on this process and compare our standard model numerical results with
those in previous work. We present the leading-order and QCD
next-to-leading-order corrected total cross sections and the
distributions of the transverse momenta of the final electron, the light
neutral Higgs boson, and jet in the minimal supersymmetric
standard model. Our results show that the
scale dependence of the leading-order cross section is obviously reduced by the
QCD next-to-leading-order corrections. The K factor of the QCD correction to the total
cross section at the large hadron-electron collider varies from $0.893$ to $1.048$ when the
factorization/renormalization scale $\mu$ goes up from $0.2 m_Z$ to
$3.8 m_Z$ in our chosen parameter space.
\end{abstract}


\vfill \eject

\baselineskip=0.32in

\renewcommand{\theequation}{\arabic{section}.\arabic{equation}}
\renewcommand{\thesection}{\Roman{section}.}
\newcommand{\nb}{\nonumber}

\newcommand{\Dir}{\kern -6.4pt\Big{/}}
\newcommand{\Dirin}{\kern -10.4pt\Big{/}\kern 4.4pt}
\newcommand{\DDir}{\kern -7.6pt\Big{/}}
\newcommand{\DGir}{\kern -6.0pt\Big{/}}

\makeatletter      
\@addtoreset{equation}{section}
\makeatother       

\section{Introduction}
\par
One of the most significant tasks for high-energy experiments is to
search for scalar Higgs
particles\cite{LHC-1,LHC-2,LHC-3,LHC-4,LHC-5}. Although the standard
model (SM)\cite{sm} has achieved impressive experimental success,
the Higgs boson, which is predicted by the SM for spontaneous
electroweak symmetry breaking, remains a mystery. Moreover,
there exists the problem of the quadratically divergent
contributions to the corrections to the Higgs boson mass, which is
the so-called naturalness problem. Alternative conceptional
difficulties, such as the hierarchy problem, the necessity of the
tuning and the nonoccurrence of gauge coupling unification at high
energies, suggest that the SM is probably the low-energy limit of a
more fundamental theory.

\par
As the most hopeful extensions of the SM, the supersymmetric (SUSY)
models can solve such problems mentioned above. The minimal
supersymmetric standard model (MSSM)\cite{mssm-1,mssm-2} is the
simplest one among all the SUSY extensions of the SM. In this model,
two Higgs doublets $H_1$ and $H_2$ give masses to up- and down-type
fermions. The Higgs sector consists of three neutral Higgs bosons,
one $CP$-odd particle ($A^0$), two $CP$-even particles ($h^0$ and
$H^0$), and a pair of charged Higgs bosons ($H^{\pm}$). However,
these Higgs bosons have not been directly explored experimentally
until now. The LEP experiments provided lower mass bounds as: for
the SM Higgs boson $m_{H^0}>114.4~GeV$ (at $95\%$ C.L.), and for the
MSSM bosons $m_{h^0}>92.8~GeV$ and $m_{A^0}>93.4~GeV$ for
$\tan\beta>0.4$ (at $95\%$ C.L.)\cite{hepdata,LEP}.

\par
Recently, a possible high-energy collider in $e^-$-$p$ collision
mode at the LHC, the large hadron-electron collider (LHeC), has been
sketched\cite{LHeC1,LHeC2}. There will exist a rich physics
program\cite{Klein}. The LHeC can be used to accurately determine
the parton dynamics and the momentum distributions of quarks and
gluons in the proton, and furthermore it may play a significant role in
the discovery and interpretation of new physics. The incoming proton
beam at the LHeC has an energy $E_p= 7~TeV$ and the energy of
incoming electron is considered as $E_e = 50 - 200~GeV$ according to
several scenarios, with the center-of-mass system energy of
$\sqrt{s}=2\sqrt{E_p E_e}\approx 1.18 - 2.37~TeV$. It seems that the
LHeC provides a cleaner environment than a hadron-hadron collider in
accessing the couplings of the Higgs boson to gauge bosons.

\par
The production channel \epehj, a neutral current (NC) process at the
LHeC, attracted the physicist's attentions. In Ref.\cite{THan} it is
pointed out that the electron reconstruction in the NC process is
superior with respect to that of the missing neutrino in the charged
current process, $e^-p \to \nu_eh^0j+X$, and the NC process has
the potential to increase the overall Higgs boson signal efficiency,
and there they studied the use of forward jet tagging as a means to
secure the observation of the Higgs boson in the $H^0 \to b \bar b$
decay mode and to significantly improve the purity of the signal.
The QCD next-to-leading-order (NLO) corrections to the SM Higgs productions of $e^-p \to
e^-H^0j+X$ and $e^-p \to \nu_eH^0j+X$ processes at the LHeC were
calculated by B. J\"ager in Ref.\cite{BJager}. Moreover, not only
does this channel provide a spectacular signature ($e^-b\bar{b}j$),
but also the lightest Higgs $h^0$ production in MSSM via vector
boson fusion with unusual visible decays is
possible\cite{future_collider}. The coupling strength of the
lightest Higgs $h^0$ with $Z^0Z^0$ is distinguished from the SM Higgs
one with an additional factor $\sin(\beta-\alpha)$, where $\beta$ is
related to the ratio of the vacuum expectation values and $\alpha$
is the mixing angle of the two CP-even Higgs states. Therefore, we
may disentangle between the SM Higgs and the light MSSM CP-even
Higgs by measuring the cross section for $e^-p \to e^- h^0 j + X$ at
the LHeC when $|\sin(\beta-\alpha)|$ is smaller than 1. Besides,
in order to find new physics it requires sufficiently precise
predictions for the new physics signals and their backgrounds with
multiple final particles which cannot be separated in
experimental data entirely. Therefore, the higher order QCD predictions for
these reactions are necessary.

\par
In this paper, we calculate the full QCD NLO corrections to the
process \epehj at the LHeC and estimate the capability of the LHeC
to access the light MSSM $CP$-even Higgs boson in the $e^-h^0j$
production. The numerical results at the leading-order (LO) are compared with
those in Ref.\cite{THan}. The paper is organized as follows: We
describe the technical details of the related LO
and QCD NLO calculations in both the SM and the MSSM in Secs. II and III, respectively. In Sec. IV we give some numerical results and
discussions about the QCD NLO corrections in the MSSM. Finally, a
short summary is given.

\vskip 5mm
\section{LO cross sections }
\par
The LO and QCD NLO calculations are carried out in 't Hooft-Feynman
gauge. The {\sc FeynArts 3.4} package\cite{fey} is adopted for generating
Feynman diagrams and subsequently converting them to corresponding
amplitudes. The {\sc FormCalc 5.4} program\cite{formloop} is applied to
reduce the amplitudes.

\par
\begin{figure*}
\begin{center}
\includegraphics[scale=1.0]{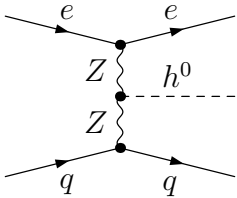}
\caption{\label{fig1} The LO Feynman diagram for the partonic
processes \eqehq ($q=u,\bar{u},d,\bar{d},c,\bar{c},s,\bar{s}$). }
\end{center}
\end{figure*}

In calculating the $e^-p \to e^-h^0j+X$ process in the MSSM, we
neglect the u-, d-, c-, s-quark masses ($m_u = m_d = m_c = m_s =
0$), and do not consider the partonic processes with incoming
(anti)bottom-quark due to the heavy (anti)bottom-quark suppression
in parton distribution functions (PDFs) of proton. That means we
involve the contributions of the following partonic processes in our
LO calculations:
\begin{eqnarray}\label{partonic process}
\label{process1} &&  e^-(p_{1})+q(p_{2})\to
e^{-}(p_{3})+h^0(p_{4})+q(p_{5}),~~
(q=u,\bar{u},d,\bar{d},c,\bar{c},s,\bar{s}),
\end{eqnarray}
where $p_{i}(i=1,...,5)~$ represent the four-momenta of the incoming
electron, partons, and the outgoing electron, $h^0$-boson and jet,
respectively. The LO Feynman diagram for the partonic processes
(\ref{partonic process}) is depicted in Fig.\ref{fig1}.

\par
The expression of the LO cross section for the partonic process
$e^-q \to e^-h^0q$ can be written in the form as
\begin{eqnarray}
\label{sigma_eq} \hat{\sigma}_{LO}(\hat{s}, e^-q \to e^-h^0q)&=&
\frac{1}{2 \hat{s}} \int \overline{\sum}|{\cal M}_{LO} |^2
d\Omega_{3},~~(q\,=\,u,\bar{u},d,\bar{d},c,\bar{c},s,\bar{s}),
\end{eqnarray}
where $\hat{s}$ is the partonic center-of-mass (c.m.) energy
squared, the summation is taken over the spins and colors of final
states, and the bar over the summation means taking average over the
intrinsic degrees of freedom of initial particles, and $d \Omega_3$
is the three-body phase-space element for the \eqehq subprocess.
${\cal M}_{LO}$ in Eq.(\ref{sigma_eq}) is the tree-level amplitude
for the partonic process \eqehq. The coupling of the SM Higgs boson
($H^0$) to the $Z^0$ pair can be expressed as $g_{HZZ}^{SM}=
\frac{ie}{c_W^2} \frac{m_W}{s_W} g^{\mu\nu}$, while the light
$CP$-even SUSY Higgs boson to the $Z^0$ pair is expressed as
$g_{hZZ}^{MSSM}= \frac{ie}{c_W^2} \frac{m_W}{s_W}
\sin(\beta-\alpha)g^{\mu\nu}$.

\par
The LO total cross section for the $e^-p \to e^{-} h^0j+X$ process
at the LHeC can be expressed as
\begin{eqnarray}\label{sigma_PP}
&& \sigma_{LO}(e^-p\to e^{-}h^0j+X)=   \nonumber \int dx
\sum_{q=u,\bar{u},d,\bar{d}}^{c,\bar{c},s,\bar{s}} \left[
G_{q/p}(x,\mu_f)\hat{\sigma}_{LO}(xs,\mu_f, e^-q \to
e^-h^0q)\right].
\end{eqnarray}
There $\mu_f$ is the factorization scale, $s$ is the total c.m.
energy squared of the electron-proton collision, $x$ describes the
four-momentum fraction of parton $q$ in an incoming proton with the
definitions of $x=\frac{p_{2}}{P}$, and $P$ is the four-momentum of
the incoming proton. $G_{q/p}$ ($q
=u,\bar{u},d,\bar{d},c,\bar{c},s,\bar{s}$) represent the PDFs of
parton $q$ in proton $p$.

\vskip 5mm
\section{ QCD NLO corrections in the MSSM}
\par
\subsection{Virtual corrections}
\par
\begin{figure*}
\begin{center}
\includegraphics[scale=1.0]{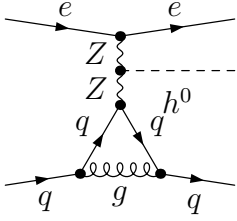}
\caption{\label{fig2a} The SM QCD vertex diagram for the partonic
process $e^-q \to e^{-}h^0q$ ($q
=\,u,\bar{u},d,\bar{d},c,\bar{c},s,\bar{s}$). }
\end{center}
\end{figure*}
\begin{figure*}
\begin{center}
\includegraphics[scale=1.0]{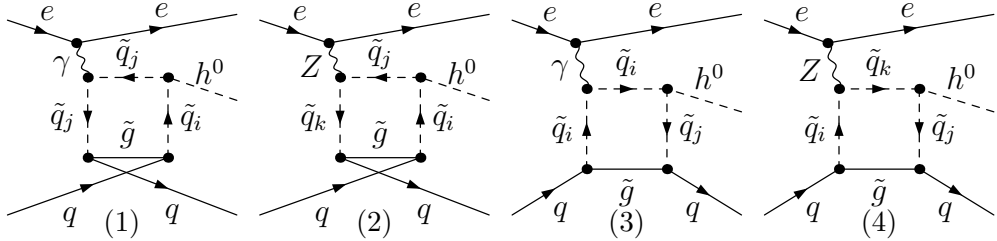}
\caption{\label{fig2b} The representative pure SUSY QCD one-loop
Feynman diagrams for the partonic process $e^-q \to e^{-}h^0q$,
where $\tilde{q} =\tilde{u},\tilde{d},\tilde{c},$ $\tilde{s}$ and
the lower indexes $i,j,k=1,2$. }
\end{center}
\end{figure*}

In order to compare the results in the MSSM with those in the SM we
present the QCD NLO calculations in both models. In the NLO
calculations, we adopt the dimensional regularization in
$D=4-2\epsilon$ dimensions to isolate the ultraviolet (UV) and
infrared (IR) singularities. The wave functions of the external
fields are renormalized under the on-shell renormalization
scheme. The virtual correction to the subprocess \eqehq involves
both soft and collinear IR singularities. In our calculation we
introduce the following counterterms for related wave functions in
the SM and the MSSM:
\begin{eqnarray}
\label{defination of renormalization constants} && \psi_{q,L}^0 =
\left(
1+\frac{1}{2}\delta Z_{q,L}^{SM(MSSM)} \right)\psi_{q,L}, \nb \\
&& \psi_{q,R}^0 = \left( 1+\frac{1}{2}\delta Z_{q,R}^{SM(MSSM)}
\right)\psi_{q,R},~~(q=u,d,c,s).
\end{eqnarray}
The wave-function renormalization constants of the massless quarks
($q=u,d,c,s$) in the SM are written as
\begin{equation} \label{CT-SM}
\delta Z_{q,L}^{SM} =\delta Z_{q,R}^{SM}= - \frac{\alpha_s}{3 \pi}
\left[\Delta_{UV}-\Delta_{IR}\right],
\end{equation}
where $\Delta_{UV}=1/\epsilon_{UV} -\gamma_E +\ln(4\pi)$ and
$\Delta_{IR}=1/\epsilon_{IR} -\gamma_E +\ln(4\pi)$. The explicit
expressions for the one-loop QCD wave-function renormalization
constants of the massless quarks ($q=u,d,c,s$) in the MSSM have the
forms as
\begin{eqnarray} \label{CT-SUSY1}
 \delta Z_{q,L}^{MSSM}  =- \frac{\alpha_s}{3 \pi}
\left[\Delta_{UV}-\Delta_{IR}\right] + \frac{2\alpha_s}{3\pi }
\left[
B_1(0,m^2_{\tilde{g}},m^2_{\tilde{q}_1})\cos^2\theta_{\tilde{q}}+
B_1(0,m^2_{\tilde{g}},m^2_{\tilde{q}_2})\sin^2\theta_{\tilde{q}}\right],
\end{eqnarray}
\begin{eqnarray}\label{CT-SUSY2}
\delta Z_{q,R}^{MSSM} = - \frac{\alpha_s}{3 \pi}
\left[\Delta_{UV}-\Delta_{IR}\right] + \frac{2\alpha_s}{3 \pi }
\left[
B_1(0,m^2_{\tilde{g}},m^2_{\tilde{q}_1})\sin^2\theta_{\tilde{q}}+
B_1(0,m^2_{\tilde{g}},m^2_{\tilde{q}_2})\cos^2\theta_{\tilde{q}}\right],
\end{eqnarray}
where the definitions for the two-point integrals are adopted from
Ref.\cite{COMS}, and $\theta_{\tilde{q}}$ is the mixing angle of
scalar quarks ($\tilde{q}_{L},~\tilde{q}_{R}$),
\begin{equation}\label{mixing angle}
\tilde{q}_{L}= \tilde{q}_{1}\cos{\theta_{\tilde{q}}} - \tilde{q}_{2}
\sin{\theta_{\tilde{q}}}, ~~~\tilde{q}_{R}= \tilde{q}_{1}
\sin{\theta_{\tilde{q}}} + \tilde{q}_{2} \cos{\theta_{\tilde{q}}}.
\end{equation}
The one-loop level Feynman diagrams include self-energy, vertex, box
(4-point) and counterterm Feynman graphs. We depict the SM QCD
vertex diagram in Fig.\ref{fig2a}, and the representative pure SUSY
QCD (pSQCD) one-loop diagrams are drawn in Fig.\ref{fig2b}.

\vskip 5mm
\subsection{Real gluon and light-(anti)quark emission corrections }
\par
The relevant real emission partonic processes can be grouped as (1)
$e^-q \to e^{-}h^0qg$, (2) $e^-g\to e^{-}h^0q\bar{q}$. There the
quark notation, $q$, represents
$u-,\bar{u}-,d-,\bar{d}-,c-,\bar{c}-,s-$ and $\bar{s}-$quark. The
real gluon/light-(anti)quark emission partonic channels (1) and (2)
at the tree-level contain soft and collinear IR singularities. After
the summation of the virtual corrections with all the real parton
emission corrections, the numerical result is soft IR-safe, while
collinear divergences still remain. It will be totally IR-safe when
we include the contributions from the collinear counterterms of the
PDFs. The IR finiteness can be verified numerically in our numerical
calculations.

\par
\begin{figure*}
\begin{center}
\includegraphics[scale=1.0]{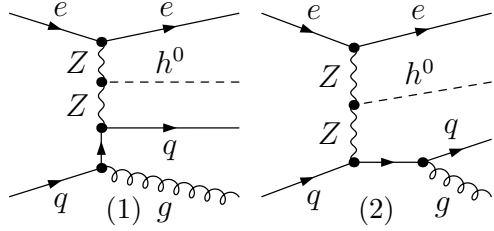}
\caption{\label{fig3a} The tree-level Feynman diagrams for the gluon
emission partonic process $e^-q \to e^{-}h^0qg$ ($q
=u,\bar{u},d,\bar{d},c,\bar{c},s,\bar{s}$). }
\end{center}
\end{figure*}
\begin{figure*}
\begin{center}
\includegraphics[scale=1.0]{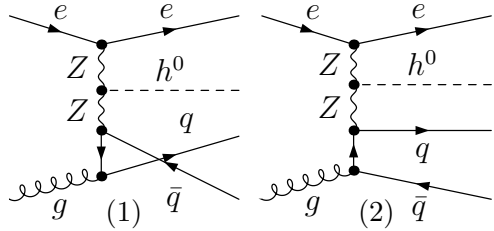}
\caption{\label{fig3b} The tree-level Feynman diagrams for the
light-quark emission partonic process $e^-g \to e^{-}h^0q\bar{q}$
($q =u,d,c,s$). }
\end{center}
\end{figure*}

The IR singularities of the real parton emission subprocesses can be
isolated by adopting the two cutoff phase-space slicing
method\cite{19}. In Figs.\ref{fig3a} and \ref{fig3b} we present
the Feynman diagrams for the real gluon emission subprocess
$e^-(p_1)q(p_2) \to e^{-}(p_3)h^0(p_4)q(p_5)g(p_6)$ and real
light-(anti)quark emission subprocess $e^-(p_1)g(p_2) \to
e^{-}(p_3)h^0(p_4)q(p_5)\bar{q}(p_6)$, respectively. In adopting
the two cutoff phase-space slicing
method we introduce an arbitrary small soft cutoff
$\delta_{s}$ to separate the $2 \to 4$ phase-space into two regions,
$E_{6}\leq\delta_{s}\sqrt{\hat{s}}/2$ (soft gluon region) and
$E_{6}>\delta_{s}\sqrt{\hat{s}}/2$ (hard gluon region), and another
cutoff $\delta_{c}$ to decompose the hard region into a hard
collinear (HC) region with $p_{2}(p_{5}) . p_6 < \delta_c \hat s/2$
and hard noncollinear ($\overline{HC}$) region with $ p_{2}(p_{5}) .
p_6 \geq \delta_c \hat s/2$.

\par
Then the cross sections for the real emission subprocesses $e^-(q,g)
\to e^{-}h^0q(g,\bar{q})$ can be written as
\begin{equation}
\label{sigmaR} \hat{\sigma}_{R}=\hat{\sigma}^{S}+\hat{\sigma}^{H}
=\hat{\sigma}^{S}+\hat{\sigma}^{HC}+\hat{\sigma}^{\overline{HC}}.
\end{equation}

\par
\section{ Numerical Results and Discussion}
\par
In our numerical calculations we take one-loop and two-loop running
$\alpha_{s}$ in the LO and NLO calculations,
respectively\cite{hepdata}. The QCD parameters are taken as $N_f=5$,
$\Lambda_5^{LO}=165~MeV$ and $\Lambda_5^{\overline{MS}}=226~MeV$. We
take the renormalization and factorization scales to be a common
value as $\mu \equiv \mu_r=\mu_f$ and choose the energy scale to be
at the $Z^0$ mass (i.e., $\mu=\mu_0=m_Z$) by default, which
characterizes the typical momentum transfer in the process \epehj.
The relevant SM parameters are taken as $m_e = 0.511~MeV$,
$m_b=4.2~GeV$, $m_t=171.2~GeV$, $m_W = 80.398~GeV$, $m_Z =
91.1876~GeV$ and $G_F=1.16637\times10^{-5}~GeV^{-2}$\cite{hepdata},
and thus we get $\alpha=1/132.34$ by adopting the relation of
$\alpha = \frac{\sqrt{2}}{\pi} G_F m_W^2 s_W^2$. We use the PDFs of
CTEQ6L1 and the CTEQ6M in the LO and NLO calculations,
respectively\cite{cteq}.

\par
The related SUSY parameters, such as the mixing angle of the MSSM
Higgs fields $\alpha$ and masses of the light $CP$-even neutral
Higgs boson, gluino, and scalar quarks, are obtained from the
FormCalc package, except otherwise stated. The input parameters for
the {\sc FormCalc} program are $M_S$, $M_2$, $A_f$, $m_{A^0}$, $\mu$ and
$\tan\beta$. There $M_Q=M_U=M_D=M_S$ and the soft trilinear
couplings for squarks $\tilde{q}$ being equal, i.e., $A_q=A_l=A_f$
are assumed, and the grand unification theory relation $M_1 =
(5/3)\tan^2 \theta_W M_2$ is adopted for simplification. In our
numerical calculation, we set $M_S=400~GeV$, $M_2=110~GeV$,
$m_{A^0}=150~GeV$, $\mu=-200~GeV$, $A_f=800~GeV$, $\tan\beta=3$, and
$m_{\tilde{g}}=230~GeV$ in default. Then we get
$\sin(\beta-\alpha)=0.9347$, $m_{\tilde{t}_1}=198.17~GeV$,
$m_{\tilde{t}_2}=579.67~GeV$,
$m_{\tilde{u}_1}=m_{\tilde{c}_1}=397.07~GeV$,
$m_{\tilde{u}_2}=m_{\tilde{c}_2}=398.76~GeV$,
$\theta_{\tilde{u}}=\theta_{\tilde{c}}=\pi/2$,
$m_{\tilde{d}_1}=m_{\tilde{s}_1}=400.62~GeV$,
$m_{\tilde{d}_2}=m_{\tilde{s}_2}=403.52~GeV$,
$\theta_{\tilde{d}}=\theta_{\tilde{s}}=0$ and $m_{h^0}=98.36~GeV$.
In the FormCalc program the radiative corrections to the MSSM Higgs
boson masses up to two-loop contributions are involved, and the
expressions related to the mass of the light neutral $CP$-even Higgs
boson in Ref.\cite{mh0} are adopted, where the input parameters of
$m_b$, $m_t$, $m_{\tilde{t}_1}$ and $m_{\tilde{t}_2}$ are necessary.

\par
The verifications for the total QCD NLO correction being independent
of the two cutoffs $\delta_s$ and $\delta_c$ are made. We calculate
the total QCD NLO corrections to the $e^-p \to e^-h^0j+X$ process in
the MSSM at the LHeC with the cutoffs $\delta_s$ running from
$10^{-5}$ to $10^{-3}$, $\delta_c = \delta_s/200$, and
$\mu=\mu_0=m_Z$. The results show that although the three-body
correction [$\Delta\sigma^{(3)}=\sigma^{V} +\sigma^{S}+\sigma^{HC}$]
and four-body correction
[$\Delta\sigma^{(4)}=\sigma^{\overline{HC}}$] depend strongly on the
cutoff $\delta_s$ ($\delta_c$), the final total QCD NLO correction
$\Delta\sigma_{NLO}$, which is the summation of the three-body and
four-body terms, i.e.,
$\Delta\sigma_{NLO}=\Delta\sigma^{(3)}+\Delta\sigma^{(4)}$ is
independent of the two cutoffs within the statistic errors. The
independence of the full QCD NLO corrections to the $e^-p \to
e^-q(q=u,\bar u) \to e^-h^0j+X$ process on the cutoffs $\delta_s$
and $\delta_c$ provides an indirect check for the correctness of the
calculations. In further numerical calculations, we fix $\delta_s=8
\times 10^{-4}$ and $\delta_c=\delta_s/200$.

\par
We made the comparison of our LO numerical results for the process
$e^-p \rightarrow e^-H^0j+X$ in the SM at the LHeC with the
corresponding results read out from Fig.2 in Ref.\cite{THan},
and find that they are coincident with each other within the
statistic errors.

\par
In the following LO and NLO numerical calculations, we adopt the
massless four-flavor scheme and put the restriction of
$p_{T}^{j}>p_{T,j}^{cut}$ on the jet transverse momentum for
one-jet events. For the two-jet events (originating from the real
corrections), we apply the jet algorithm in the definition of the
tagged hard jet with $R=1$, i.e., if final state two partons
satisfy $\sqrt{\Delta \eta^2 + \Delta \phi^2} < 1$ (where $\Delta
\eta$ and $\Delta \phi$ are the differences of rapidity and
azimuthal angle between the two jets), we merge them into a single
jet. We use the so-called "inclusive" scheme and keep events with
one or two jets. We require that there is one jet with $p_{T}^{j}
> p_{T,j}^{cut}$, and set $p_{T,j}^{cut}=30~GeV$ by default in
following calculations. Furthermore, to reduce the background of
the Higgs signals, we require the final electron with the following
cuts
\begin{equation}
p_T^e>30~GeV,\quad |\eta^e|<5.
\end{equation}

\par
\begin{figure}[htbp]
\includegraphics[scale=0.8]{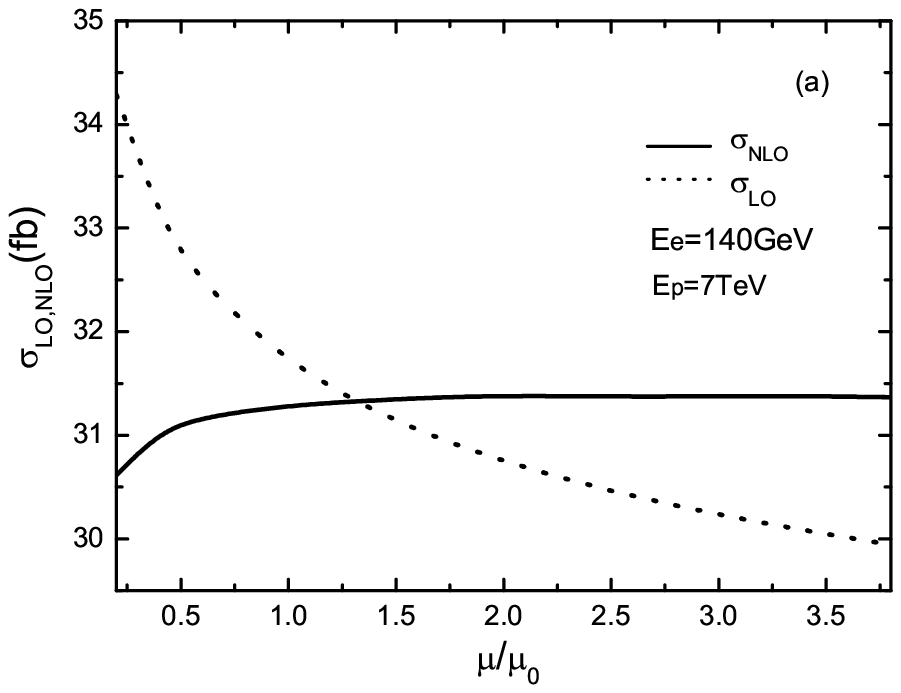}
\includegraphics[scale=0.8]{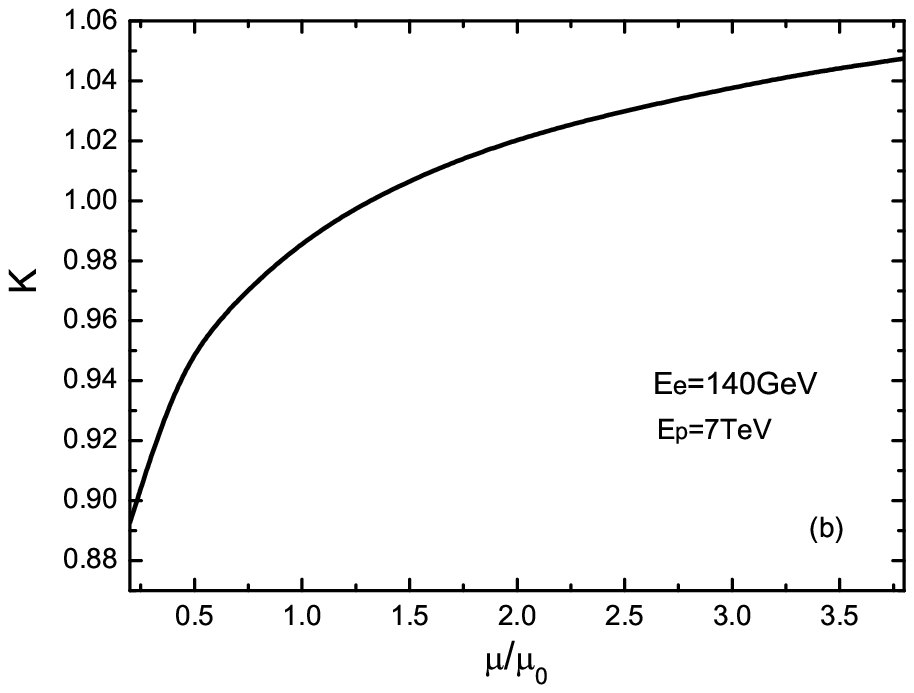}
\hspace{0in}%
\caption{\label{fig5} (a) The dependence of the LO and QCD NLO
corrected total cross sections for the \epehj process on the
renormalization/factorization scale $\mu=\mu_r=\mu_f$ in the MSSM,
where we take $E_p=7~TeV$ and $E_e=140~GeV$. (b) The corresponding
K factor of Fig.\ref{fig5}(a) versus the energy scale $\mu$ (where
we define $K=\frac{\sigma_{NLO}}{\sigma_{LO}}$). }
\end{figure}

We plot the dependence of the LO and QCD NLO corrected total cross
sections for the \epehj process in the MSSM on the
renormalization/factorization scale $\mu$ in Fig.\ref{fig5}(a). The
corresponding K factor defined as $K =
\frac{\sigma_{NLO}}{\sigma_{LO}}$, versus the energy scale is
presented in Fig.\ref{fig5}(b). Figure\ref{fig5}(a) shows that the LO
curve is obviously dependent on the energy scale $\mu$, although
there only the factorization scale is involved in the convolution
with the PDFs of the initial parton. If we define the scale
uncertainty parameter as $\eta =|\frac{\sigma(\mu_1)
-\sigma(\mu_2)} {\sigma(\mu_0)}|$ in the scale range of
$[\mu_1=\frac{1}{3} \mu_0, \mu_2=3 \mu_0]$, from Fig.\ref{fig5}(a)
we can get the uncertainty parameters for the LO and QCD NLO
corrected cross sections having the values as $\eta_{LO}=10.35\%$
and $\eta_{NLO}=1.58\%$, respectively. It is obvious that the
dependence of the LO total cross section on the energy scale is
significantly reduced by the QCD NLO corrections. We can read out
from Fig.\ref{fig5}(b) that when the energy scale varies from $0.2
m_Z$ to $3.8 m_Z$, the value of the K factor increases from $0.893$
to $1.048$. In the following, we choose $\mu=\mu_0$ except otherwise
stated.

\par
\begin{figure}[htbp]
\includegraphics[scale=0.8]{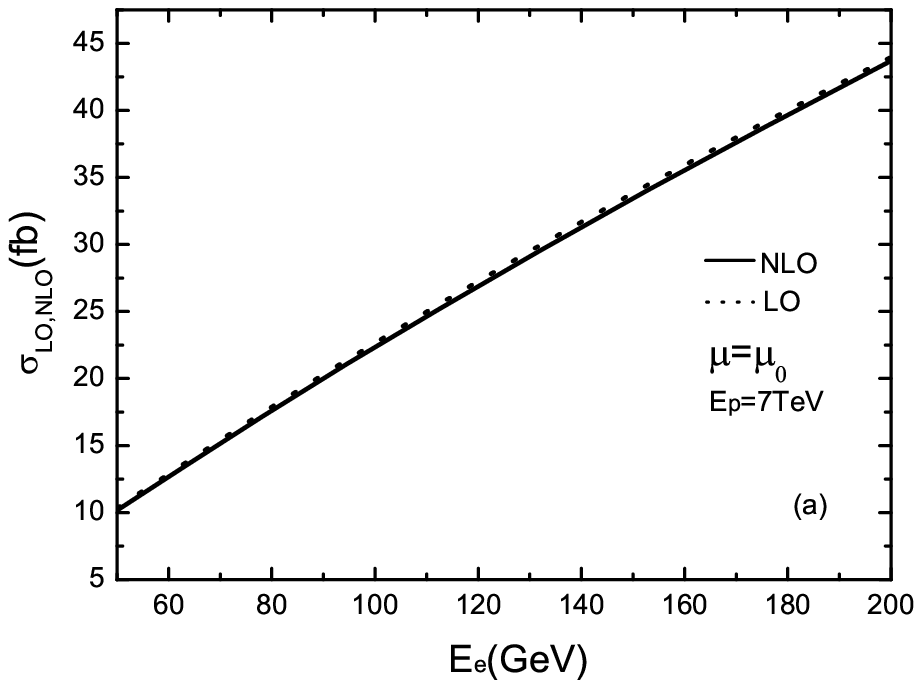}
\includegraphics[scale=0.8]{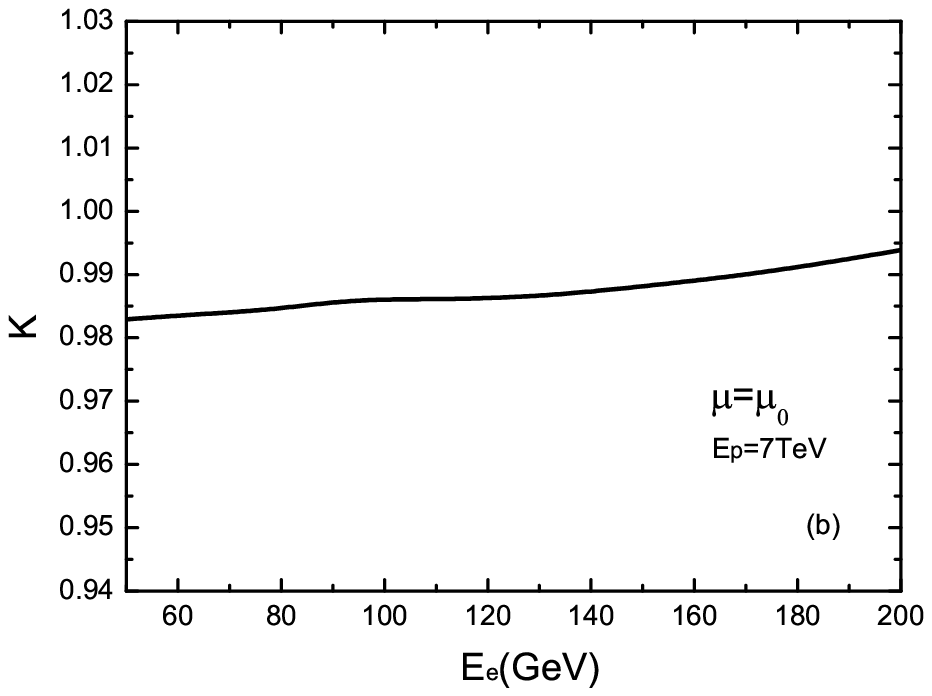}
\includegraphics[scale=0.8]{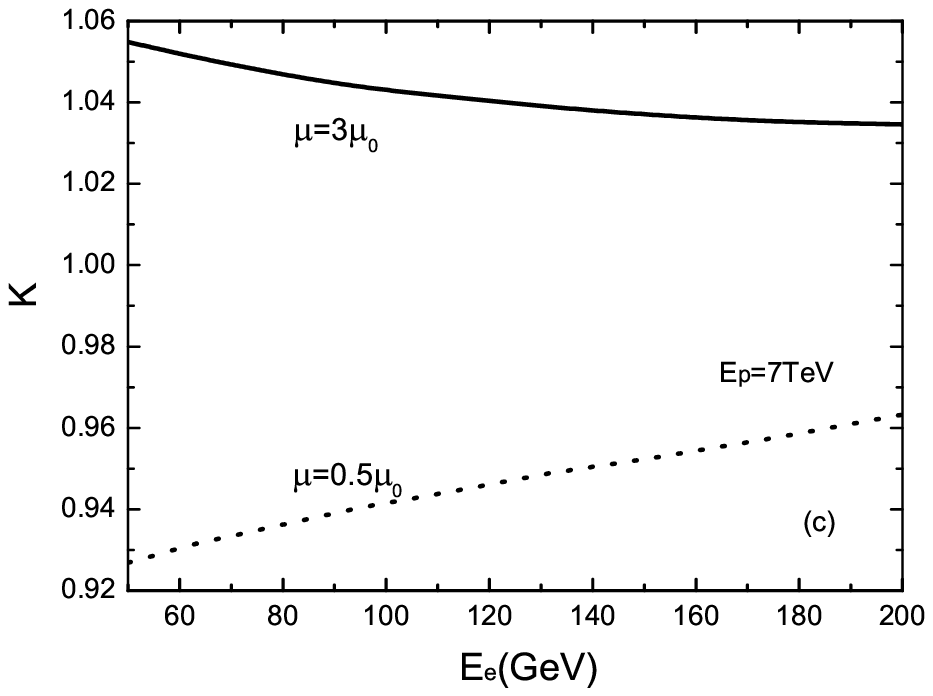}
\hspace{0in}%
\caption{\label{fig6} (a) The dependence of the LO and QCD NLO
corrected total cross sections for the \epehj process in the MSSM on
the incoming electron beam energy $E_e$ in the MSSM, where we take
$\mu=\mu_0$, $E_p=7~TeV$, and $E_e=140~GeV$. (b) The corresponding
K factor ($K=\frac{\sigma_{NLO}}{\sigma_{LO}}$) versus the incoming
electron beam energy $E_e$. (c) The K factor versus the incoming
electron beam energy $E_e$ with $\mu=0.5\mu_0$ and $\mu=3\mu_0$,
respectively. }
\end{figure}

We plot the LO and QCD NLO corrected total cross sections for the
\epehj process in the MSSM as a function of the incoming electron
beam energy $E_e$ running from $50~GeV$ to $200~GeV$ in
Fig.\ref{fig6}(a), that corresponds to the c.m. colliding energy
range of $\sqrt{s} \approx 1.18-2.37~TeV$. The corresponding
K factors are depicted as a function of the incoming electron beam
energy $E_e$ in Fig.\ref{fig6}(b). In Fig.\ref{fig6}(a) the
full line is for the QCD NLO corrected total cross section for the
\epehj process, and the dotted line for the LO cross section. We can
see from Figs.\ref{fig6}(a) and \ref{fig6}(b) that the QCD NLO corrections reduce
slightly the LO total cross sections for the process \epehj in the
plotted incoming electron beam energy range, and the production rate
increases with $E_e$. In Fig.\ref{fig6}(c) we depict the K factor
versus electron beam energy $E_e$, the energy scales $\mu$ being
$0.5\mu_0$ and $3\mu_0$ separately. We can see from
Fig.\ref{fig6}(c) that the K-factor uncertainty, $\Delta K=
K(\mu=3\mu_0)-K(\mu=0.5\mu_0)$, ranges from $12.79\%$ to $7.13\%$
when $E_e$ goes up from $50~GeV$ to $200~GeV$.

\par
\begin{figure}[htbp]
\includegraphics[scale=0.8]{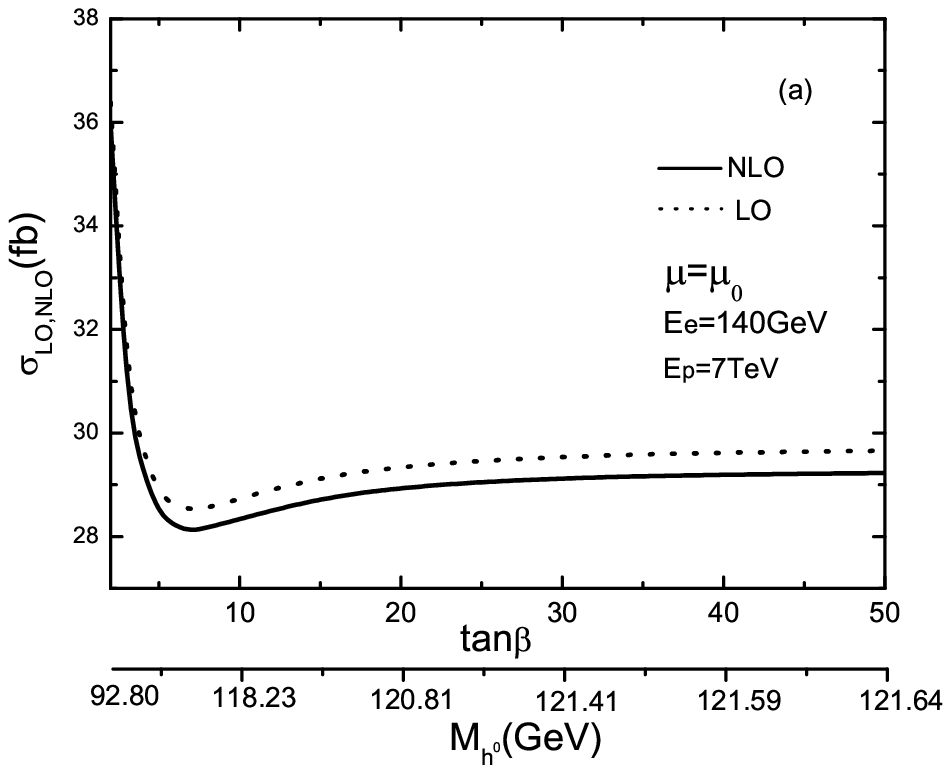}
\includegraphics[scale=0.8]{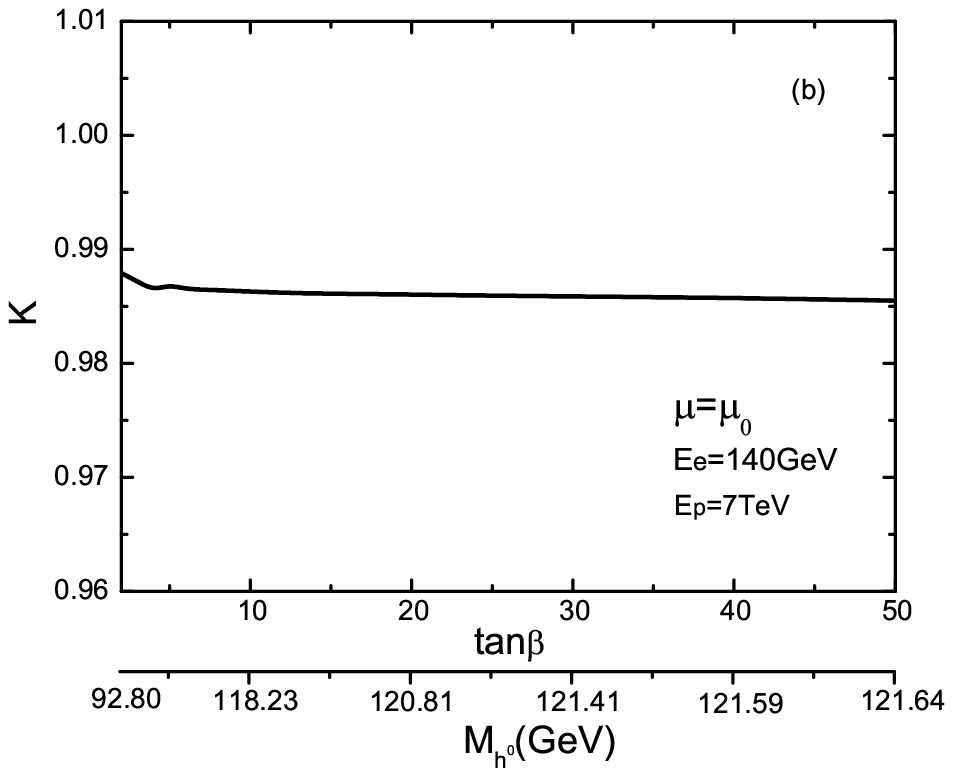}
\includegraphics[scale=0.8]{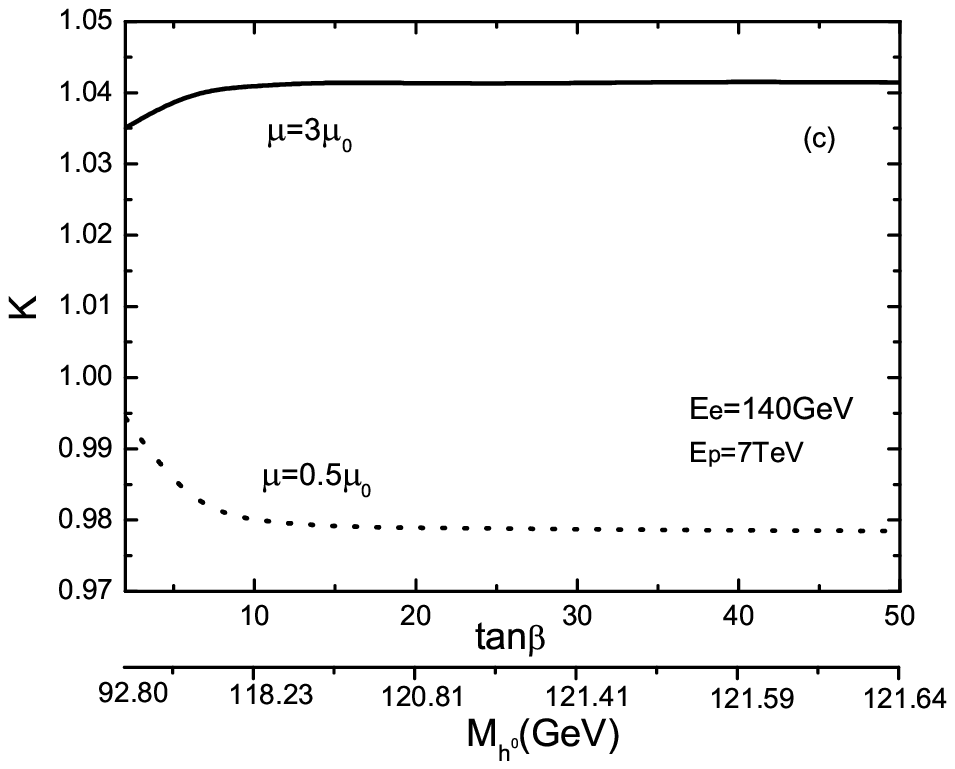}
\hspace{0in}%
\caption{\label{fig7} (a) The LO and QCD NLO corrected total cross
sections for the \epehj process as a function of $\tan\beta$
($m_{A^0}$ fixed) and the corresponding mass of the light $CP$-even
neutral Higgs boson $m_{h^0}$ in the MSSM, where we take
$\mu=\mu_0$, $E_p=7~TeV$, and $E_e=140~GeV$. (b) The corresponding
K factor ($K=\frac{\sigma_{NLO}}{\sigma_{LO}}$) versus $\tan\beta$
and $m_{h^0}$. (c) The K factor versus $\tan\beta$ and $m_{h^0}$
with $\mu=0.5\mu_0$ and $\mu=3\mu_0$, respectively. }
\end{figure}

The curves for the LO and QCD NLO corrected cross sections for the
process \epehj as a function of $\tan\beta$ are drawn in
Fig.\ref{fig7}(a), where the corresponding values of $m_{h^0}$ are
also shown on the x axis in Figs.\ref{fig7}(a) and \ref{fig7}(b).
The values of $m_{A^0}$ and of the other parameters are those given
above. In Fig.\ref{fig7}(a), we can see that both curves go down
rapidly in the region of $2 < \tan\beta < 6$ ($85.52~GeV<
m_{h^0}=113.14 ~GeV$). Then the curves go up slowly after the values
reach their corresponding minimal values at position around
$\tan\beta \sim 7.5$. The relevant K-factor
$(K=\sigma_{NLO}/\sigma_{LO})$ versus $\tan\beta$ (and $m_{h^0}$) is
plotted in Fig.\ref{fig7}(b). The K factor generally has a constant
value of about $0.99$. We further depict two curves for the K factors
with $\mu=0.5\mu_0$ and $\mu=3\mu_0$ separately, as a function of
$\tan\beta$ (and $m_{h^0}$) in Fig.\ref{fig7}(c). We can read out
from Fig.\ref{fig7}(c) that the K-factor uncertainty due to the
scale $\mu$, defined as $\Delta K = K(\mu=3\mu_0)-K(\mu=0.5\mu_0)$,
is in the range from $4.06\%$ to $6.29\%$ when $\tan\beta$
($m_{h^0}$) varies from 2 ($92.80~GeV$) to 50 ($121.64~GeV$).

\par
For the comparison of the results for the processes $e^-p \to
e^-H^0j+X$ in the SM and \epehj in the MSSM at the LHeC, we read
out the data in the MSSM from Fig.\ref{fig7}(a) at the positions
of $\tan\beta=3,7,18,38$ respectively, and list these results
together with the corresponding SM ones in Table \ref{tab2}. All
SM parameters, including the mass of the SM Higgs boson, have the
same values in both the SM and the MSSM calculations. The relative
difference between the cross sections in both models, is defined
as $\delta_{NLO} \equiv
\frac{\sigma_{NLO}^{MSSM}-\sigma_{NLO}^{SM}} {\sigma_{NLO}^{SM}}
\times 100\%$. These numbers are obtained by adopting the values
of the renormalization/factorization scale $\mu$ and the input
SUSY parameters mentioned above. From the table we can see that
the relative difference, $\delta_{NLO}$, between the cross
sections in both models can reach the value of $-10.57\%$ up to
the QCD NLO, when we take $\tan\beta=3$ for the MSSM.
\begin{table}
\begin{center}
\begin{tabular}{|c|c|c|c|c|c|c|c|}
\hline  $tan\beta$ & $m_{h^0}(m_{H^0})$ & $\sigma_{LO}^{MSSM}(fb)$ &
$\sigma_{NLO}^{MSSM}(fb)$
&$K_{MSSM}$ & $\sigma_{LO}^{SM}(fb)$ & $\sigma_{NLO}^{SM}(fb)$ & $\delta_{NLO}$ \\
\hline   3  & 98.36GeV  & 31.68(2)  & 31.29(9) & 0.988 &  36.26(3) & 34.99(9) & -10.57\%    \\
\hline   7  & 115.09GeV & 28.52(2)  & 28.14(9) & 0.987 &  31.43(3) & 31.02(9) & -9.28\%    \\
\hline   18 & 120.65GeV & 29.24(2)  & 28.85(9) & 0.987 &  29.99(3) & 29.60(9) & -2.53\%    \\
\hline   38 & 121.57GeV & 29.57(2)  & 29.18(9) & 0.987 &  29.76(3) & 28.72(9) & 1.60\%    \\
\hline
\end{tabular}
\end{center}
\begin{center}
\begin{minipage}{15cm}
\caption{\label{tab2} The numerical results of the
$\sigma_{LO}^{MSSM}$, $\sigma_{NLO}^{MSSM}$ for
$\tan\beta=3,7,18,38$ obtained from Fig.\ref{fig7}(a), and the
corresponding SM results of the $\sigma_{LO,NLO}^{SM}$ of the
process $e^-p \to e^-H^0j+X$ are listed in the table, where we take
the same SM parameters and the mass of the Higgs boson
($m_{h^0}=m_{H^0}$) in both the SM and the MSSM calculations.
$\delta_{NLO}$ is defined as $
\frac{\sigma_{NLO}^{MSSM}-\sigma_{NLO}^{SM}} {\sigma_{NLO}^{SM}}
\times 100\%$.  }
\end{minipage}
\end{center}
\end{table}

\par
\begin{figure}[htbp]
\includegraphics[scale=0.7]{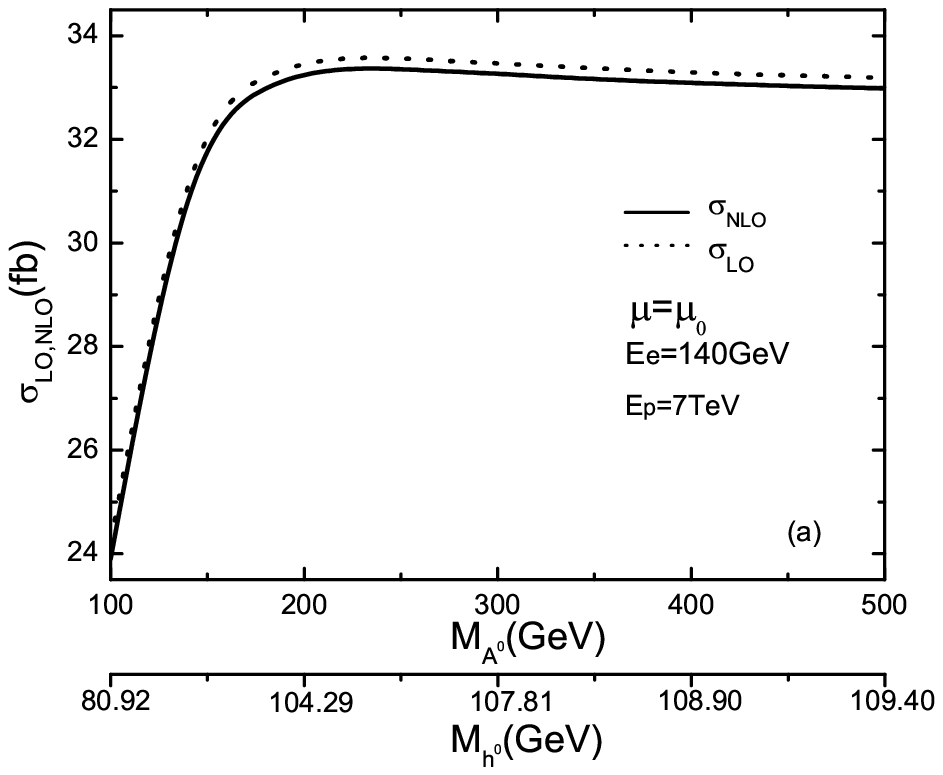}
\includegraphics[scale=0.7]{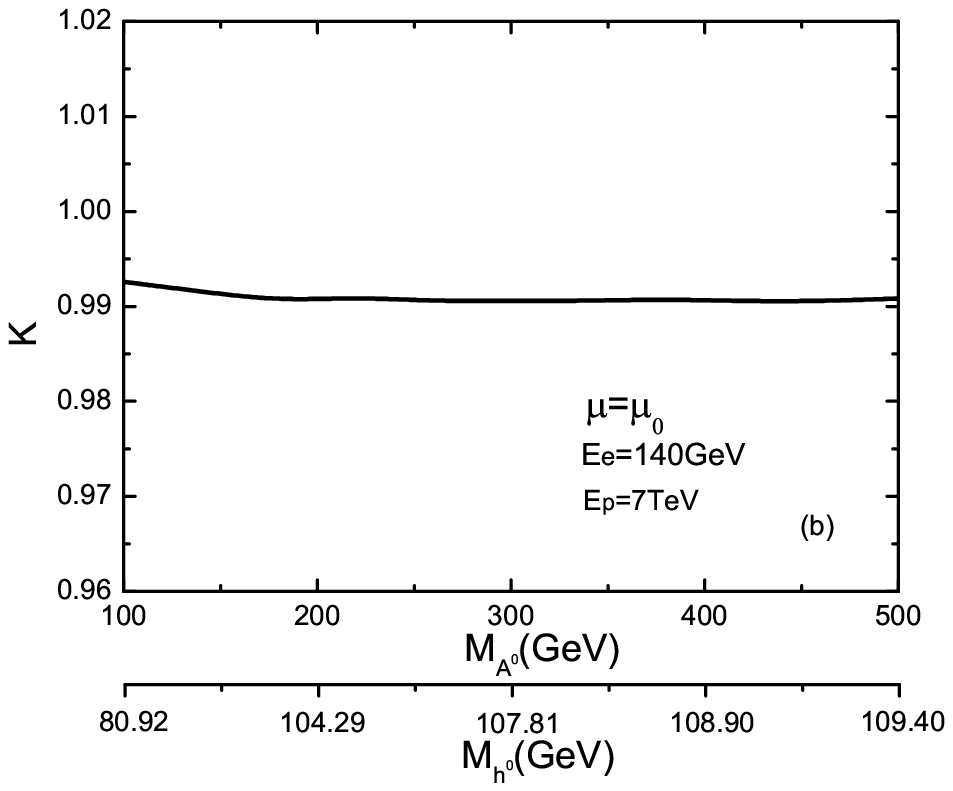}
\hspace{0in}%
\caption{\label{fig8} (a) The LO and QCD NLO corrected total cross
sections for the \epehj process as a function of the masses of the
CP-odd Higgs boson $A^0$ and the light $CP$-even neutral Higgs boson
$h^0$ ($\tan\beta$ fixed) in the MSSM, where we take $\mu=\mu_0$,
$E_p=7~TeV$ and $E_e=140~GeV$. (b) The corresponding K factor$
(K=\sigma_{NLO}/\sigma_{LO})$ as a function of $m_{A^0}$ and
$m_{h^0}$.  }
\end{figure}

In Fig.\ref{fig8}(a) we depict the LO and QCD NLO corrected cross
sections for the process \epehj as a function of mass $m_{A^0}$ (and
$m_{h^0}$). As we know, the light CP-even Higgs boson mass $m_{h^0}$
depends on the CP-odd Higgs boson mass $m_{A^0}$, when the other
related SUSY and SM input parameters are fixed. The values of
$m_{h^0}$ corresponding to different $m_{A^0}$ values are also shown
on the x axis in Figs.\ref{fig8}(a) and \ref{fig8}(b). In Fig.\ref{fig8}(a), we see
that the cross sections increase rapidly in the range of $100~GeV <
m_{A^0} < 180~GeV$ (It corresponds to the range of $80.92~GeV <
m_{h^0} < 106.02~GeV$). After reaching their maximal values at
position of $m_{A^0}=220~GeV$, the LO and QCD NLO corrected cross
sections decrease gently. The corresponding K factor versus
$m_{A^0}$ (and $m_{h^0}$) is displayed in Fig.\ref{fig8}(b). The
K factor seems to be stable and has the value around $0.99$. We can
see that when we fix the energy scale $\mu=\mu_0$, the QCD NLO
corrections in the MSSM generally reduce the LO cross section by
about $1\%$, while the pure SUSY QCD (pSQCD) NLO contributions are
negligibly small and the relative pSQCD corrections have the values
less than $0.01\%$. But as seen earlier [Fig.\ref{fig5}(b)], when
$\mu=0.2 m_Z$ ($3.8 m_Z$) the QCD NLO relative correction reaches
$-10.7\%$ ($4.8 \%$).

\par
\begin{figure}[htbp]
\includegraphics[scale=0.8]{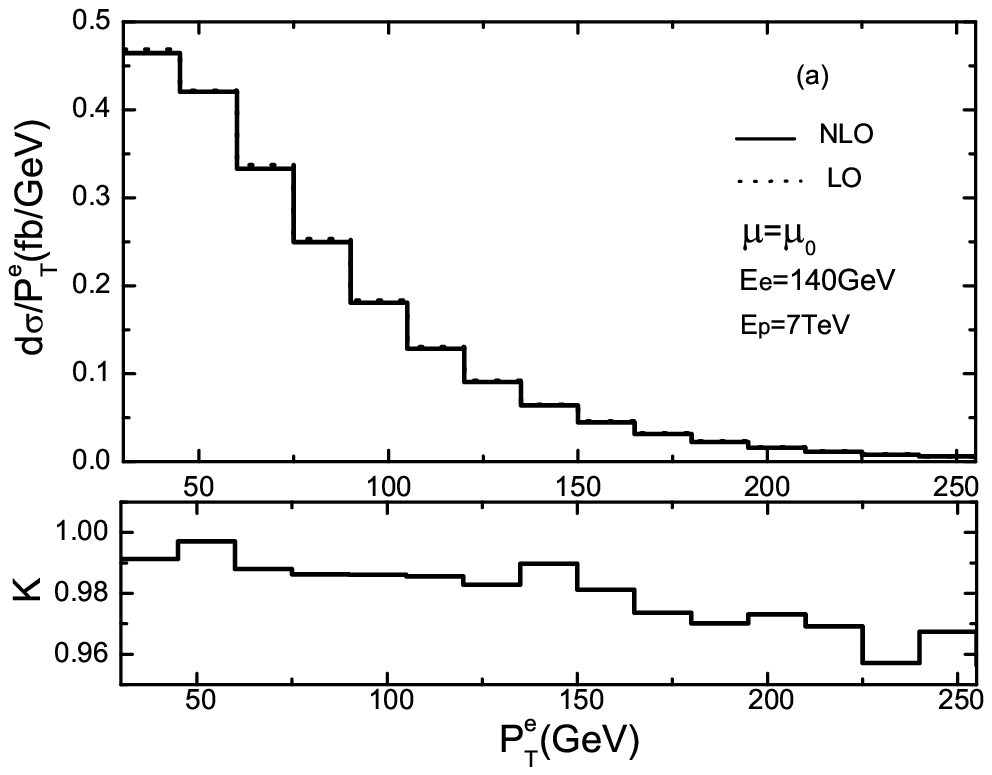}
\includegraphics[scale=0.8]{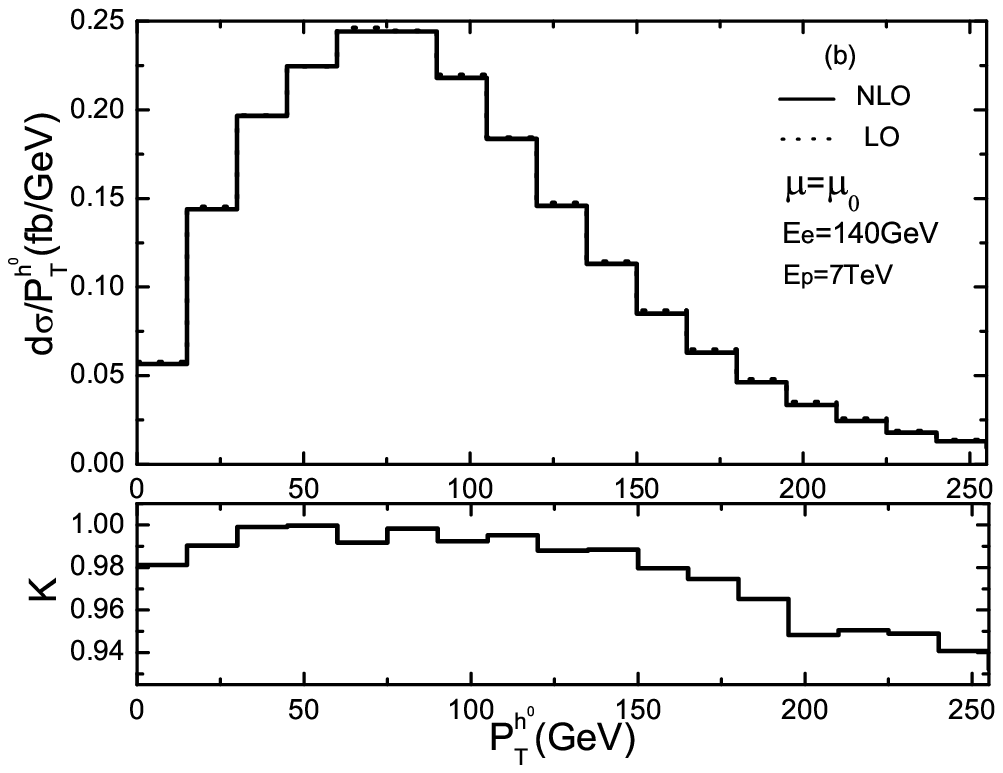}
\includegraphics[scale=0.8]{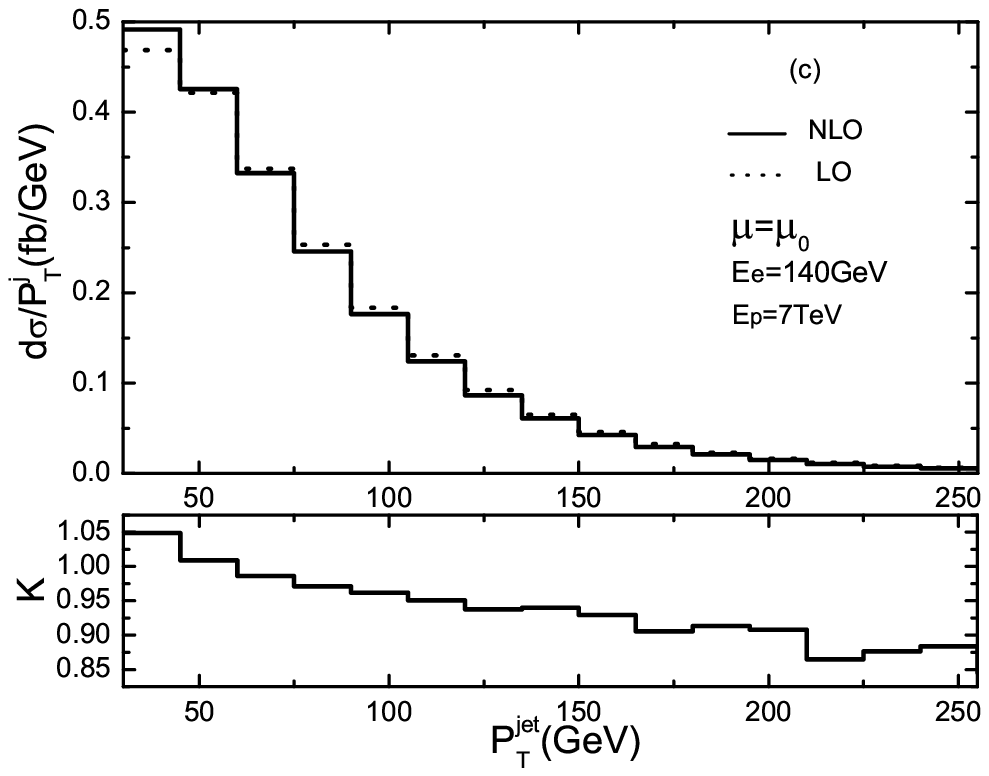}
\hspace{0in}%
\caption{\label{fig9} (a) The LO and QCD NLO corrected differential
cross sections $\frac{d \sigma}{d p_T^{e}}$ and the corresponding
K factor $\left(K = \frac{d \sigma_{NLO}}{d p_T^{e}}/\frac{d
\sigma_{LO}}{d p_T^{e}}\right)$ for the process \epehj. (b) The LO
and QCD NLO corrected differential cross sections $\frac{d \sigma}{d
p_T^{h}}$ and the corresponding K factor $\left(K = \frac{d
\sigma_{NLO}}{d p_T^{h}}/\frac{d \sigma_{LO}}{d p_T^{h}}\right)$ for
the process \epehj. (c) The LO and QCD NLO corrected differential
cross sections $\frac{d \sigma}{d p_T^{jet}}$ and the corresponding
K factor $\left(K = \frac{d \sigma_{NLO}}{d p_T^{jet}}/\frac{d
\sigma_{LO}}{d p_T^{jet}}\right)$ for the process \epehj.   }
\end{figure}

The distributions of the transverse momenta of the final particles
at the LO and up to the QCD NLO, and their corresponding K factors
for the process \epehj are depicted in Figs.\ref{fig9}(a,b,c), where
we define $K = \frac{d \sigma_{NLO}}{d p_T}/\frac{d \sigma_{LO}}{d
p_T}$. In Figs.\ref{fig9}(a), (b) and (c), the distributions of
transverse momenta and K factors are for the final electron, the
light $CP$-even neutral Higgs boson and jet, respectively. We can
find that there is no obvious distortion induced by the QCD NLO
corrections for the $p_T^{e}$ and $p_T^{h^0}$ distributions, while
the shape distortion for the $p_T^{jet}$ distribution is not
negligible since the K factor of the $p_T^{jet}$ distribution varies
in the range of $0.865 < K_{p_T^{jet}} <1.049$.

\par
\section{Summary}
\par
In this paper we calculate the full QCD NLO corrections to the light
$CP$-even neutral Higgs boson production associated with an electron
and a jet in the MSSM at the possible CERN LHeC. We investigate the uncertainty of the integrated
cross sections induced by the factorization/renormalization scale,
and present the LO and QCD NLO corrected total cross sections and
the distributions of the transverse momenta of final particles. By
adopting the definition of the scale uncertainty parameter in the
scale range of $[\mu_1=\frac{1}{3} \mu_0, \mu_2=3 \mu_0]$ as $\eta
\equiv |\frac{\sigma(\mu_1) -\sigma(\mu_2)} {\sigma(\mu_0)}|$, we
obtain the scale uncertainty parameters for the LO and NLO corrected
cross sections are $10.35\%$ and $1.58\%$, respectively. It is clear
that the scale dependence of the LO cross section is obviously
improved by the QCD NLO corrections. We find that there is no
obvious distortion induced by the QCD NLO corrections for the
$p_T^{e^-}$ and $p_T^{h^0}$ distributions, and the K factor of the
QCD correction to the total cross section at the LHeC varies from
$0.893$ to $1.048$ when the factorization/renormalization scale
$\mu$ goes up from $0.2 m_Z$ to $3.8 m_Z$ in our chosen parameter
space.

\vskip 5mm
\par
\noindent{\large\bf Acknowledgments:} This work was supported in
part by the National Natural Science Foundation of China
(No.10875112, No.11075150, No.10875112), and the Specialized
Research Fund for the Doctoral Program of Higher Education
(No.20093402110030).

\end{document}